\documentclass[%
superscriptaddress,
prx-quantum,
nofootinbib,
notitlepage
]{revtex4-2}
\usepackage{graphicx}
\usepackage{amssymb,amsmath,amsthm,amsfonts}
\usepackage{thmtools}
\usepackage{mathtools}
\mathtoolsset{centercolon}
\usepackage{thm-restate} 
\usepackage{mathtools}
\usepackage{amsmath}
\usepackage{amssymb}
\usepackage{comment}
\usepackage{placeins}
\usepackage{multirow}
\usepackage{qcircuit}
\usepackage{dcolumn}
\usepackage{makecell}
\usepackage{enumerate}
\usepackage[shortlabels]{enumitem}
\usepackage[ruled,lined]{algorithm2e}
\usepackage[caption=false]{subfig}
\usepackage[colorlinks]{hyperref}

\newcommand{\B}[1]{\textbf{#1}}

\newcommand{\Google}{\affiliation{
Google Research, Venice, CA 90291, United States}}

\begin{document}

\title{A comment on ``Factoring integers with sublinear resources on a superconducting quantum processor"}

\date{\today}

\author{Tanuj Khattar}
\email[Corresponding author: ]{tanujkhattar@google.com}
\Google

\author{Noureldin Yosri}
\email[Corresponding author: ]{noureldinyosri@google.com}
\affiliation{Google, Dublin, Ireland}

\begin{abstract}

Quantum computing has the potential to revolutionize cryptography by breaking classical public-key cryptography schemes, such as RSA and Diffie-Hellman. However, breaking the widely used 2048-bit RSA using Shor's quantum factoring algorithm is expected to require millions of noisy physical qubits and is well beyond the capabilities of present day quantum computers. A recent proposal by Yan et. al. tries to improve the widely debated Schnorr’s lattice-based integer factorization algorithm using a quantum optimizer (QAOA), and further claim that one can break RSA 2048 using only 372 qubits. In this work, we present an open-source implementation of the algorithm proposed by Yan et. al. and show that, even if we had a perfect quantum optimizer (instead of a heuristic like QAOA), the proposed claims don't hold true. Specifically, our implementation shows that the claimed sublinear lattice dimension for the Hybrid quantum+classical version of Schnorr's algorithm successfully factors integers only up to 70 bits and fails to find enough factoring relations for random 80 bit integers and beyond. We further hope that our implementation serves as a playground for the community to easily test other hybrid quantum + classical integer factorization algorithm ideas using lattice based reductions.

\end{abstract}

\maketitle
All code and data used in the paper is available at \href{https://github.com/google-research/google-research/tree/master/factoring_sqif}{github.com/google-research/google-research/factoring\_sqif}

{
  \hypersetup{linkcolor=black}
  \tableofcontents
}

\section{Introduction}
Integer factorization is a central problem to cryptography. While the exact complexity class of the problem is not known, the lack of a polynomial time algorithm suggests that it might not be in $\mathcal{P}$. The perceived difficulty of the problem made it the center piece of many cryptography algorithms, including the RSA \cite{rsa78} cryptosystem, and thus forms a foundation of modern information security.

In 1994, Peter Shor \cite{Shor1994Algorithm} showed that integer factorization can be solved efficiently on a quantum computer, which was a historic milestone that greatly increased interest in quantum computing. Today it's estimated that the number of physical qubits needed to break 2048-bit RSA is roughly 20 million \cite{Gidney_2021}, which is far beyond the capabilities of today's quantum computers.

In classical cryptography, ideas for reducing integer factorization to solving Closest Vector Problem (CVP) on a lattice were first introduced by an influential proposal by Claus-Peter Schnorr in 1991 \cite{Sch91}. The approach has been seriously explored (eg: \cite{vera2010note}), and successful applications were found outside the realm of factoring. One of the primary bottlenecks is that the method requires solving CVP on large lattices, with lattice dimension that scale polynomially in $n=\log_{2}(N)$, the bitsize of the number $N$ to be factored. In 2021, a follow up work by Schnorr \cite{schnorr2021fast} claimed to reduce the required dimension of the lattice to $\mathcal{O}(\frac{n}{\log_2{n}})$, and thus ``destroy the RSA cryptosystem". However, the claims in the paper are highly debated \cite{crypto:88601}, \cite{twitter:inf_0_} and most experts believe the lattice dimension required is much larger than what is claimed in \cite{schnorr2021fast}. Léo Ducas also published an open-source implementation \cite{SchnorrGate2021} of the proposed algorithm and the experimental results confirmed that the proposed claims did not hold true.

A recent work by Yan et. al \cite{yan2022factoring} claims to improve the drawbacks of \cite{schnorr2021fast} by using a quantum heuristic optimizer for solving CVP on a lattice, instead of a classical heuristic optimizer. They reuse the weak claim of \cite{schnorr2021fast}, that solving CVP on a lattice of dimension $\mathcal{O}(\frac{n}{\log_2{n}})$ would be enough, and thus arrive at a quantum algorithm for factoring that requires ``sublinear" resources. They further extend this claim to report resource estimate of only $37$ qubits for breaking RSA-128 and $372$ qubits for breaking RSA-2048, which took the world by surprise \cite{Castelvecchi2023} \cite{schneier2023Blog}.

The paper by Yan et. al \cite{yan2022factoring} led some people to believe that there’s been a decisive advance on how to factor huge integers, and thereby break the RSA cryptosystem, using a near-term quantum computer \cite{Castelvecchi2023} \cite{schneier2023Blog}. From a technical point of view, \cite{grebnev2023pitfalls} gives a critique of the assumptions and points out inconsistencies in the paper.

Many experts have argued \cite{cargo_cult} that the heuristic quantum optimization algorithm - QAOA \cite{farhi2014quantum}, used by \cite{yan2022factoring} is not expected to provide any meaningful speedup for solving CVP on a lattice and thus the algorithm proposed by \cite{yan2022factoring} does not lead to meaningful improvements over classical Schnorr's algorithm. Follow-up works \cite{hegade2023digitizedcounterdiabatic} have also appeared which try to replace the QAOA based quantum optimizer with other quantum heuristics that might perform better. However, all of this misses a more fundamental flaw in  the proposal, i.e., even if a perfect quantum optimizer existed for solving CVP on a lattice; the classical assumption of using a sublinear lattice itself is flawed and hence the proposal is not expected to work!

The goal of this paper is to
\begin{itemize}
    \item Provide a working implementation, with configurable hyper parameters, of a hybrid algorithm that combines Schnorr's classical lattice based algorithm for integer factorization with a quantum optimizer for solving CVP on a lattice.
    \item Use our implementation to refute the claims of \cite{yan2022factoring} and show that using a sublinear prime lattice with a brute-force based perfect quantum optimizer succeeds for factoring RSA style random integers upto 80-bits but fails beyond that. A QAOA based quantum optimizer performs much worse, as expected.
\end{itemize}

We hope our open-source implementation serves as a playground for the community to empirically test the various unproved ideas in classical and quantum algorithms for lattice based integer factorization. For example, to the best of our knowledge, this is the first implementation of Schnorr's algorithm that succeeds in factoring upto 80-bit integers with a sublinear lattice of dimension $26$. 

\section{Implementation of Sublinear Resource Quantum Integer Factorization (SQIF) Algorithm}
Most of the pitfalls of the SQIF algorithm proposed by Yan et. al. \cite{yan2022factoring} were covered by \cite{grebnev2023pitfalls}. Here, we mainly aim to focus on the pitfall that the claims for requiring sublinear quantum resources are based on Schnorr's algorithm \cite{schnorr2021fast}, whose efficiency and probability of success are unknown.

The classical Schnorr's algorithm \cite{schnorr2021fast} is not known to scale efficiently, in particular \cite{SchnorrGate2021} did an experimental study that showed that the algorithm fails for 40-bit numbers.
The SQIF algorithm by Yan et. al. \cite{yan2022factoring} builds on top of the classical Schnorr's algorithm, with minor modifications to the lattice construction scheme, and proposes the use of a quantum computer to find higher quality (closer) solutions to the CVP problem on a lattice. This leads to a hybrid ``classical+quantum" algorithm such that

\begin{enumerate}[I.]
    \item\label{sqif-classical} Classical: Use the claims from \cite{schnorr2021fast} to reduce the integer factorization problem to solving the (approximate) Closest Vector Problem (CVP), which is NP-hard \cite{Dinur2003}, on a lattice of dimension $m=\mathcal{O}(\frac{n}{\log_2{n}})$. 

    \item\label{sqif-quantum} Quantum: Find an approximate solution to CVP on a lattice using a classical heuristic method, like Babai's algorithm \cite{Babai1986}, and further improve the classical solution by considering all neighbouring $2^m$ lattice points in the $m$-dimensional unit length cube around the classical solution. The second optimization is done by finding ground states of an all to all connected Ising Hamiltonian via a heuristic quantum optimizer like QAOA \cite{farhi2014quantum}.

\end{enumerate}

While experts of quantum computing have raised concerns \cite{cargo_cult} about QAOA's performance on finding ground states of the proposed Hamiltonian (i.e. step \ref{sqif-quantum}), we empirically show that the classical reductions of step \ref{sqif-classical} itself are flawed and thus a hypothetical perfect quantum optimizer for step \ref{sqif-quantum} would also not help provide any meaningful improvement over classical Schnorr's algorithm.   

Our observations are consistent with the experimental study conducted by Léo Ducas \cite{SchnorrGate2021} to test the claims of Schnorr's factoring algorithm \cite{schnorr2021fast}, where he showed that using a sublinear sized lattice would not be sufficient for obtaining enough factoring relations and that the required lattice dimension would be much larger than what is claimed in \cite{schnorr2021fast}.

In algorithm \ref{algo:FactoringQAOA}, we present an overview of the SQIF factorization algorithm and we highlight all hyper-parameters of the algorithm, for a discussion of their effect and values see Appendix \ref{appendix:hyperparameters}. For the details of the algorithm we refer the reader to our implementation \cite{FactoringSQIF}, or to the original proposal by Yan et. al\cite{yan2022factoring}

\begin{algorithm}
    \caption{Quantum Integer Factorization using Schnorr's algorithm as described in \cite{yan2022factoring}.} \label{algo:FactoringQAOA}
    \SetKwInOut{Input}{input}
    \Input{N}
    \Input{\textbf{lattice\_parameter}}
    \Input{\textbf{precision\_parameter}}
    \Input{\textbf{smoothness\_bound}}
    \Input{\textbf{num\_energy\_states\_samples}}
    \Input{\textbf{qaoa\_depth}}
    \Input{\textbf{method}}
    \Begin{
        $\textit{n} \gets \textit{lattice\_dimension}(N, \textbf{\textit{lattice\_parameter}})$\;
        $sr\_pairs \gets \emptyset$\\
        \While{$|sr\_pairs| < \textit{num\_required\_sr\_pairs}(n, \textbf{\textit{smoothness\_bound}})$}{
            $B, t \gets \textit{construct\_random\_closest\_vector\_problem}(n, \textbf{\textit{precision\_parameter}})$ \\
            $b_{opt}, B, D, \mu \gets \textit{approximate\_cvp\_using\_babai}(B, t) $ \\
            $H \gets \textit{construct\_hamiltonian}(t, b_{opt}, D, \mu)$\\
            \eIf{$\textit{method} = \textit{QAOA}$} {
                $states \gets \textit{sample\_low\_energy\_states\_using\_qaoa}$(H,\textbf{\textit{qaoa\_depth}})\\ 
            } {
                $states \gets \textit{brute\_force\_find\_low\_energy\_states}$(H,\textbf{\textit{num\_energy\_states\_samples}})\\           
            }
            $\textit{lattice\_vectors} \gets \textit{lattice\_vectors\_from\_states}(states, B)$ \\
            $sr\_pairs \gets sr\_pairs \cup \textit{sr\_pairs\_from\_lattice\_vectors}(\textit{lattice\_vectors}, \textbf{\textit{smoothness\_bound}})$\\
        }
        $A \gets \textit{difference\_of\_exponents\_matrix}(sr\_pairs, N, \textbf{\textit{smoothness\_bound}})$\\
        $\textit{factors} \gets \emptyset$\\
        \For{$z \in \textit{null\_space}(A \textit{ mod } 2)$} {
            $e \gets \frac{1}{2} Az$\\
            $X \gets \prod_{e_i > 0} p_i^{e_i}$\\            
            $Y \gets \prod_{e_i < 0} p_i^{-e_i}$\\
            $\textit{factors} \gets \textit{factors} \cup \{gcd(X + Y, N), gcd(|X - Y|, N)\}$
        }
        \Return{factors}
    }
\end{algorithm}

\section{Results}
Our code can be found in \cite{FactoringSQIF}. We ran our code on a Google Cloud Platform (GCP) a2-highgpu-1g virtual machine. Our code optimizes its performance by moving linear algebra and vectorizable operations to the GPU.

The largest number tested in \cite{yan2022factoring} was 261980999226229 which has 48-bits. We ran our code against random numbers with 63, 70, 80, 90, 100, 120, 128 bits. As Table. \ref{tab:results} shows our implementation succeeds in factoring numbers up to 80-bits long and then fails to factor larger numbers.

Note that the claim on page 5 of \cite{yan2022factoring} that 372 qubits to break RSA-2048 means that we need to use a lattice parameter of 2 for large numbers. Thus in our work when using a lattice parameter of 1 fails we try a lattice parameter of 2. This works for the 70-bit case and then fails at 80-bit onwards. Note that we kept doing this up 100 bits. For 120 and 128 bit numbers we only tried lattice parameter of 1 since a lattice parameter of 2 would have been infeasible to brute force.

\begin{table}[!h]
    \centering
    \begin{tabular}{|c|c|c|c|c|c|c|c|}
        \hline
        \# bits & N & \makecell{lattice \\ parameter} & \# qubits & \# SR-Pairs & method & success/fail & \makecell{\# iterations \\ to success}\\
        \hline
        \B{40} & \B{624911573291} & \B{1} & \B{11} & \B{247} & \B{qaoa} & \B{success} & \B{40}\\
        \hline
        \B{40} & \B{624911573291} & \B{1} & \B{11} & \B{247} & \B{brute force} & \B{success} & \B{40}\\
        \hline
        \B{48} & \B{261980999226229} & \B{1} & \B{12} & \B{291} & \B{qaoa} & \B{success} & \B{127} \\
        \hline
        \B{48} & \B{261980999226229} & \B{1} & \B{12} & \B{291} & \B{brute force} & \B{success} & \B{118} \\
        \hline
        63 & 2393864445846808531 & 1 & 15 & 165  & qaoa & fail & \\
        \hline
        \B{63} & \B{2393864445846808531} & \B{1} & \B{15} & \B{452} & \B{brute force} & \B{success} & \B{359}\\
        \hline
        70 & 700821480830487125167 & 1 & 17 & 45 & brute force & fail &\\
        \hline
        70 & 700821480830487125167 & 2 & 23 & 479 & qaoa & fail & \\
        \hline
        \B{70} & \B{700821480830487125167} & \B{2} & \B{23} & \B{1064} & \B{brute force} & \B{success} & \B{242}\\
        \hline
        80 & 675789769078847752141081 & 2 & 26 & 1295 & brute force & fail &\\
        \hline
        90 & 928497021444492107802357067 & 1 & 15 & 1 & brute force & fail &\\
        \hline
        90 & 928497021444492107802357067 & 2 & 30 & 446 & brute force & fail &\\
        \hline
        100 & 729097431295829382764936159407 & 1 & 14 & 0 & brute force & fail &\\
        \hline
        100 & 729097431295829382764936159407 & 2 & 28 & 33 & brute force & fail &\\
        \hline
        120 & 925141703449007503130714828237701463 & 1 & 17 & 0 & brute force & fail &\\
        \hline
        128 & 275538060341916784483102145290705042411 & 1 & 18 & 0 & brute force & fail &\\
        \hline
        \hline
    \end{tabular}
    \caption{Factorization Results}
    \label{tab:results}
\end{table}

\section{Conclusion}
The Schnorr's algorithm with a sub-linear lattice size fails to scale. Using heuristic quantum optimizers to solve CVP on a lattice in the hope of speeding up the process of finding SR pairs isn't fruitful in this case since even if we brute force all possibilities the algorithm still fails.

\newpage
\bibliographystyle{apsrev4-2}
\bibliography{references}

\appendix

\section{Hyperparameters} \label{appendix:hyperparameters}

\subsection{Lattice Parameter and Lattice Dimension}
In \cite{yan2022factoring} the function for computing the lattice dimension is given in multiple slightly different variations. We believe that Equation. \ref{eq:lattice_parameter} captures these where $\textit{lattice\_parameter (l)}$ is a hyperparameter, $N$ is the integer to be factored and $m$ is the lattice dimension.

\begin{align}
    n = \lfloor \log_2{N} \rceil \\
    m = \lfloor \frac{l * n}{\lfloor \log_2{n \rceil}} \rfloor    \label{eq:lattice_parameter}    
\end{align}

In \cite{yan2022factoring} two values were used for the hyperparameter $l \in \{1, 2\}$. In general, a larger value of $\textit{lattice\_parameter} (l)$ leads to a larger lattice dimension, which corresponds to a higher probability of finding enough good SR pairs (thus leading to a successful factorization) at the cost of increasing the space and time complexity of the algorithm.

\subsection{Precision parameter for lattice construction}
A CVP consists of a matrix $B$ of basis vectors that define a lattice and a target vector $t$. The precision parameter ($c$) affects the last row of $B$ and the final coordinate of $t$ as given by Equation. \ref{eq:cvp}

\begin{align}\label{eq:cvp}
B_{n,c} = \begin{bmatrix} 
    f(1) & 0 & \dots & 0\\
    0 & f(2) & \dots & 0\\
    \vdots & \vdots & \ddots & \vdots\\
    0 & 0 & \dots & f(n) \\
    \lfloor10^c\ln{p_1}\rceil & \lfloor10^c\ln{p_2}\rceil & \dots & \lfloor10^c\ln{p_n}\rceil 
    \end{bmatrix}
t = \begin{bmatrix}
    0 \\
    \vdots \\
    10^c \ln{N}
\end{bmatrix}
\end{align}

where the function $f(i)$ for $i = 1, ..., n$ is a random permutation of $(\lceil 1/2 \rceil, \lceil 2/2 \rceil, ..., \lceil n/2 \rceil)$.

The determinant of the lattice is positively correlated with precision parameter $c$. It is conjectured that probability of finding vectors close to the target vector should be positively correlated with $c$, but it's not clear to us that why this should be true. In our work we used $c = 4$

\subsection{Smoothness bound $B_2$}

A $B$-smooth number is a positive integer such that the largest prime that divides it is $\leq B$. A $B$-smooth relation (SR) pair is a pair of integers $(u, v)$ such that $u, v, \text{and} |u - vN|$ are all $B$-smooth. In Schnorr's algorithm, by construction, every point on the lattice can be mapped to a pair of integers $(u, v)$ s.t. both $u$ and $v$ are $p_m$-smooth, however not all of them are $B_2$ SR pairs, where $B_2$ is a hyperparameter.

In general, $B_2$ affects the algorithms as:
1) Higher the $B_2$ bound, the easier it is to find SR-pairs $(u, v)$ that satisfy the constraint that $|u - v * N|$ is $B_2$-smooth. Thus it increases the probability of finding valid SR-pairs.
2) However, an increase in $B_2$ also leads to an increase in the requirement for number of SR-pairs to sample, which increases the overall runtime of the algorithm. 

To factorize $N$, it is sufficient to find enough SR-pairs. It is typically sufficient that the number of such SR-pairs is a few more than smoothness bound $B_2$. In \cite{yan2022factoring} they suggest that $B_2 \sim 2m^2$.

\subsection{Energy states specific hyperparameters}
In our work we implemented two versions of the algorithm. The first (brute forces) computes the energy of all of the  $2^m$ states and returns the \textbf{num\_energy\_states\_samples} lowest enerergy states (we set $\textbf{num\_energy\_states\_samples} = 2^{\text{min}(15, m)}$). The second is the QAOA heuristic which constructs the circuit described in \cite{yan2022factoring} with depth \textbf{qaoa\_depth}. In \cite{yan2022factoring}  the depth was $\in [1, 3]$. In our work we used depth $= 2$.

\end{document}